\documentclass[pra,twocolumn,showpacs,preprintnumbers,amsmath,amssymb]{revtex4}
\usepackage{graphicx}% Include figure files
\usepackage{dcolumn}% Align table columns on decimal point
\usepackage{bm}% bold math
\usepackage{latexsym,epsfig}

\newcommand{\beq}{\begin{equation}}
\newcommand{\eeq}{\end{equation}}
\newcommand{\bea}{\begin{eqnarray}}
\newcommand{\eea}{\end{eqnarray}}
\newcommand{\ben}{\begin{eqnarray*}}
\newcommand{\een}{\end{eqnarray*}}
\newcommand{\be}{\begin{enumerate}}
\newcommand{\ee}{\end{enumerate}}
\newcommand{\bfig}{\begin{figure}}
\newcommand{\efig}{\end{figure}}

\newcommand{\ba}{\begin{align}}
\newcommand{\ea}{\end{align}}

\newcommand{\D}{\displaystyle}

\newcommand{\la}{\langle}
\newcommand{\ra}{\rangle}

\newcommand{\ocdw}{O_{\text{CDW}}}
\newcommand{\ocellcdw}{O^{\text{cell}}_{\text{CDW}}}

\begin{document}
%\baselineskip 24pt
% \preprint{Version 4.6}
\title{Supersolid and solitonic phases in one-dimensional Extended Bose-Hubbard model}
\author{Tapan Mishra}
\email{tapan@iiap.res.in} \affiliation{ Indian Institute of
Astrophysics, II Block, Kormangala, Bangalore, 560 034, India.}
\author{Ramesh V. Pai}
\email{rvpai@unigoa.ac.in} \affiliation{ Department of Physics, Goa
University, Taleigao Plateau, Goa 403 206, India. }
\author{S. Ramanan}
\email{suna@cts.iisc.ernet.in}
\affiliation{Centre for High Energy Physics,
Indian Institute of Science, Bangalore\ 560012, India}
\author{Meetu Sethi Luthra}
\email{meetu@iiap.res.in} \altaffiliation [permanent address
]{Bhaskaracharya College of Applied Sciences, Phase-I,
Sector-2,Dwarka,Delhi,110075, India.} \affiliation{Indian Institute
of Astrophysics, II Block, Kormangala,  Bangalore,  560 034, India.}

\author{B. P. Das}
\email{das@iiap.res.in} \affiliation{Indian Institute of
Astrophysics, II Block, Kormangala, Bangalore, 560 034, India.}

\date{\today}

\begin{abstract}
We report our findings on quantum phase transitions in cold bosonic
atoms in a one dimensional optical lattice using the finite size
density matrix renormalization group method in the framework of the
extended Bose-Hubbard model. We consider wide
ranges of values for the filling factors and the nearest neighbor
interactions. At commensurate fillings, we obtain two different
types of charge density wave phases and a Mott insulator phase.
However, departure from commensurate fillings yield the exotic
supersolid phase where both the crystalline and the superfluid
orders coexist. In addition, we obtain signatures for solitary waves
and also superfluidity.
\end{abstract}

%\pacs{75.40.Gb, 75.10.Jm, 75.30.Ds, 75.40.Mg}
\pacs{03.75.Lm, 05.10.Cc, 05.30.Jp} \keywords{Suggested keywords}

\maketitle

%%%%%%%%%%%%%%%%%%%%%%%%%%%%%%%%%%%%%%%%%%%%%%%%%%%%%%%%%%%%%%%%%%%%
\section{INTRODUCTION}
\label{sect:intro} The supersolid phase, first reported in
$^4\text{He}$~\cite{penrose},  is characterized by the coexistence
of the superfluid and crystalline orders. This phase has been
predicted in several bosonic lattice systems
~\cite{batrouniprl,kedar,sengupta,arun,wessel,scarola}, however,
there has been no
unambiguous observation of this phase so far. Kim {\it et al} had reported its observation in solid $^4\text{He}$~\cite{chan}, but a number of studies
disagree with this claim~\cite{ref1,ref2,ref3}.

In recent years, the advances in the manipulation of cold bosonic
atoms in the optical lattices have opened up a new route to
investigate quantum phase transitions~\cite{bloch,blochreview}. This
approach has many advantages over the conventional solid state
techniques, such as for example, flexibility in controlling the parameters and  the
dimension of the lattice by tuning the laser intensity. A system of
cold bosonic atoms in an optical lattice can be adequately described
by the Bose-Hubbard model~\cite{fisher,jaksch}. However, if the
atoms possess long range interactions due to the presence of
magnetic dipole moments, for example, then they could exhibit a
number of different novel phases. In  particular, the existence of
such interactions could result in the supersolid
phase~\cite{batrouniprl,scarola,mishrass,mathey}. The fairly recent
observation of the Bose-Einstein condensation  of $^{52}\text{Cr}$
atoms~\cite{pfau}, which have large magnetic dipole moments could
ultimately lead to the observation of this unusual phase.

In this context, we re-investigate the system of bosonic atoms with
the long range interaction using the extended Bose-Hubbard model
given by
 \bea \label{eq:ham}
H&=&-t\sum_{<i,j>}(a_{i}^{\dagger}a_{j}+H.c)\nonumber\\
&& \mbox{} +\frac{U}{2}\sum_{i} n_{i}(n_{i}-1)+V\sum_{<i,j>}n_in_j.
\eea
Here $t$ is the hopping amplitude between nearest neighboring
sites $<i,j>$. $a_i^{\dagger}(a_i)$ is the bosonic creation
(annihilation) operator obeying the Bosonic commutation relation
$[a_i^\dagger,a_j]=\delta_{i,j}$ and $n_i=a_i^{\dagger}a_i$ is the
number operator. $U$ and $V$ are the onsite and the nearest
neighbor interactions, respectively. We rescale in unit of the hopping amplitude,
$t$, setting $t= 1$, making the Hamiltonian and other quantities dimensionless.

In the absence of long range interactions, the model given in
Eq.~\ref{eq:ham} reduces to the Bose-Hubbard model which exhibits a
superfluid (SF) to Mott insulator (MI) transition at integer
densities of bosons~\cite{fisher}. However, for non-integer
densities, the system remains in the superfluid phase which is
compressible and gapless. The Mott insulator phase, however, has
finite gap and is incompressible. The extended Bose-Hubbard model,
given in Eq.~\ref{eq:ham}, has been studied earlier using different
methods~\cite{kashurnikov,batrouni95,niyaz,batrouniprl} including
DMRG~\cite{pai,kuhner,whiteprb}. The inclusion of the nearest
neighbor interaction gives rise to the charge density wave (CDW)
phase for integer and half integer
densities~\cite{kashurnikov,batrouni95,niyaz,batrouniprl,kuhner,pai,whiteprb}
that has finite gap, finite CDW order parameter, vanishing
compressibility and a peak in the density structure function at
momentum $q=\pi$. In the CDW phase the bosons occupy alternate sites
and the unoccupied ones being empty. For example, when the density
$\rho=1/2$, the distribution of bosons has a
$\mid{1~0~1~0~1~0~\cdots}\ra$ structure while for $\rho=1$ it is
$\mid{2~0~2~0~2~0~\cdots}\ra$. To distinguish between these two CDW
ground states, the former is referred to as CDW-I and the latter,
CDW-II~\cite{batrouniprl}. This model has been studied recently
using quantum monte-carlo~\cite{batrouniprl} resulting in the
prediction of the supersolid phase when the density of the bosons is
no longer commensurate. We re-investigate this model by departing
from both half and integer filling for large and intermediate onsite
interaction strengths and obtain the phase diagram using the finite
size density matrix renormalization group(FS-DMRG)
method~\cite{white,schollwock} and throw more light on the
supersolid and solitonic phases.

This paper is organized as follows. In Sec.~\ref{sect:meth}, we will
discuss the method of our calculation using FS-DMRG. The results
with discussion are presented in Sec.~\ref{sect:res_disc} followed
by our conclusions in Sec.~\ref{sect:concl}.

%%%%%%%%%%%%%%%%%%%%%%%%%%%%%%%%%%%%%%%%%%%%%%%%%%%%%%%%%%%%%%%%%%%
\section{METHOD OF CALCULATION}
\label{sect:meth} To obtain the ground state wave function and the
energy for the system of $N$ bosons on a lattice of length $L$,
interacting via an on-site and a nearest neighbor interaction, we
use the FS-DMRG method with open boundary
conditions~\cite{white,schollwock}. This method is best suited for
one dimensional problems and has been widely used to study the
Bose-Hubbard model~\cite{pai,kuhner,whiteprb,schollwock,laura}. We
have considered four bosonic states per site and the weights of the
states neglected in the density matrix formed from the left or right
blocks are less than $10^{-6}$~\cite{pai}. In order to improve the
convergence, at the end of each DMRG step, we use the finite-size
sweeping procedure given in~\cite{white,pai}. Using the ground state
wave function $|\psi_{LN} \rangle$ and energy $E_L(N)$, where $N$
refers to the number of bosons and $L$, the length of the lattice,
we calculate the following physical quantities and use them to
identify the different phases. The chemical potential $\mu$ of the
system having density $\rho=N/L$ is given by \beq \mu =
\D\frac{\delta E_L(N)}{\delta N} \label{eq:mu} \eeq and the gapped
and gapless phases are distinguished from the behavior of $\rho$ as
a function of $\mu$~\cite{ramanan}. The compressibility $\kappa$,
which is finite for the SF phase, is calculated using the relation
\beq
    \kappa =\D\frac{\delta \rho}{\delta \mu}.
\label{eq:kappa}
\eeq
The on-site local number density $\la n_i \ra$, defined as,
\beq
    \la n_i \ra= \langle\psi_{LN}|n_i|\psi_{LN} \rangle,
    \label{eq:ni}
\eeq
gives information about the density distribution of different
phases and finally the existence of the CDW phase is confirmed by
calculating its order parameter:
 \beq
    \ocdw=\frac{1}{L}\sum_i(-1)^i\la n_i \ra.
\label{eq:ocdw}
\eeq

When the ground state is a CDW, FS-DMRG calculation with open
boundary leads to an artificial node in the density distribution at
the center due to left-right symmetry. We circumvent this problem by
working with odd number of sites. In our calculations, we start with
five sites instead of usual choice of four sites and increase the
length up to $L=101$ adding two sites in each DMRG
iteration~\cite{pai}. After reaching the desired length $L = 101$,
we vary the number of atoms $N$ from $26$ to $125$ to scan a wide
range of densities~\cite{ramanan}. In this work we consider, two
different values of the onsite interaction strengths: $U=5$ and $10$
and vary the nearest neighbor interaction strength $V$ from $0$ to
$U$. The choice of $U$ is guided by an earlier work~\cite{pai} where
a direct MI to CDW-II transition for $U=10$ and a MI to SF to CDW-II
for $U=5$ were observed as $V$ is varied at a density $\rho=1$. In
this work, we extend this calculation to a wider range of densities
and obtain a richer phase diagram consisting of supersolid and
solitonic phases in addition to SF, CDW-I and CDW-II. We begin our
discussion for $U=10$ and later comment on our results for $U=5$.

%%%%%%%%%%%%%%%%%%%%%%%%%%%%%%%%%%%%%%%%%%%%%%%%%%%%%%%%%%%%%%%%%%%
\section{RESULTS AND DISCUSSION}
\label{sect:res_disc} It is well known that the Bose-Hubbard model
(Eq.~\ref{eq:ham}, with $V=0$) has a superfluid ground state when
density $\rho$ is not an integer and exhibits a quantum phase
transition from the superfluid to the Mott insulator for integer
densities~\cite{fisher} at a critical value of onsite interaction
$U_C$ that depends on $\rho$. (For example, $U_C\sim 3.4$ for
$\rho=1$~\cite{pai,kuhner}.) The Mott insulator has finite gap and
zero compressibility while the superfluid is gapless and
compressible. In the presence of a finite nearest neighbor
interaction $V$, an additional insulator phase, CDW, appears at
commensurate densities. As noted in~\cite{pai,kuhner}, a CDW-I
occurs at $\rho=1/2$ and at $\rho=1$, depending on the value of $V$
either a MI or a CDW-II appears. Since we are dealing with only an
on-site and a nearest neighbor interaction, the commensurate
densities for model in Eq.~\ref{eq:ham} are integers and half
integers. We begin by studying the possible phases at commensurate
densities, before we investigate the phases at incommensurate
densities.

\begin{figure}[ht]
  \centering
\includegraphics[width = 3.4in, angle = 0, clip = true]
{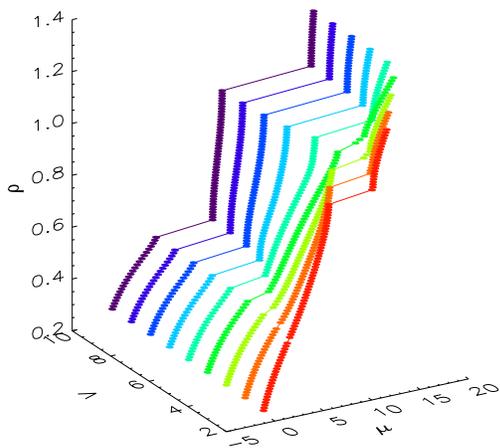}
    \caption{(Color online)The density $\rho$ as a function of the chemical potential $\mu$ for different values of $V$.
    The plateaus at the
commensurate fillings indicate the existence of finite gap in the
system.}
    \label{fig:fig1}
\end{figure}

\begin{figure}[ht]
  \centering
\includegraphics[width = 2.4in,height=3in, angle = 0, clip = true]
{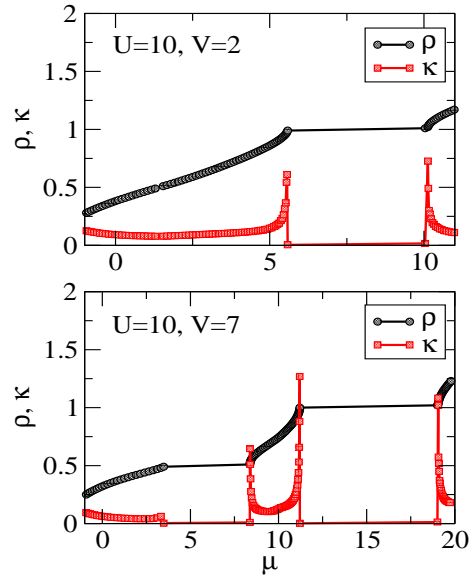}
    \caption{(Color online)Variation of $\kappa$ and $\rho$ with $\mu$ for $V=2$
(top panel) and $V=7$ (bottom panel). The plateau
regions have zero compressibility while it is finite elsewhere.}
    \label{fig:fig2}
\end{figure}

The gapped phases are easily obtained from the dependence of the
density $\rho$ and the compressibility $\kappa$ on the chemical
potential $\mu$. Figure~\ref{fig:fig1} shows the dependence of
$\rho$ on $\mu$ for a fixed value of $U=10$ and $V$ ranging between
$2$ and $10$. The gapped phases appear as plateaus with the gap
equal to the width of the plateau, i.e., $\mu^+ -\mu^-$, where
$\mu^+$ and $\mu^-$, respectively, are the values of the chemical
potential at the upper and lower knee of the plateau. For small
values of $V$, Fig.~\ref{fig:fig1} has only one plateau at $\rho=1$.
However, as we increases $V$ an additional plateau appears at
$\rho=1/2$. We calculate the compressibility using
Eq.~\ref{eq:kappa} and is also given as a function of $\mu$ in
Fig.~\ref{fig:fig2} for two generic values of $V$. The smaller
value, $V = 2$, has just one plateau at $\rho=1$ while $V= 7$, has
two, at densities $\rho=1/2$ and $1$. As expected, the
compressibility is zero over the range of $\mu$ values where the
plateaus occur, while it is finite elsewhere. The incompressible
insulator and compressible superfluid regions can be separated out
by picking up $\mu^+$ and $\mu^-$ and plotting them in the $\mu - V$
plane. From Figs.~\ref{fig:fig1} and~\ref{fig:fig2}, we see that,
(i) a gapped phase occurs at $\rho=1/2$ for $V \gtrsim 2.8$, (ii)
for $\rho=1$, the gap remains finite for all values of $V$ and (iii)
the gap is zero for other values of $\rho$.

\begin{figure}[ht]
  \centering
\includegraphics[width = 3.4in, angle = 0, clip = true]
{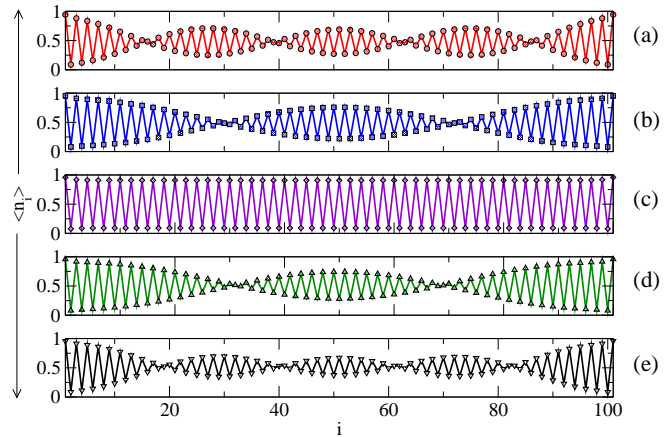}
    \caption{(Color online)Local density $\la n_i \ra$ as a function of lattice sites
$i$. Panels (a) and (b) shows the solitonic signature when $\rho<
1/2$. (c) shows the signature of CDW-I where every alternate site is
occupied with one boson for $\rho=1/2$. Panels (d) and (e)  shows
modulation of the CDW-I phase for $\rho>1/2$ and are once again in
the solitonic phase.}
    \label{fig:fig3}
\end{figure}

The nature of the compressible and incompressible phases can further
be understood from the local density distribution $\langle n_i
\rangle$ and the charge density wave order parameter $\ocdw$ given
by Eqs.~\ref{eq:ni} and~\ref{eq:ocdw}. The variation of local
density as a function of the lattice sites are given in
Figs.~\ref{fig:fig3} and~\ref{fig:fig5} for densities around
$\rho=1/2$ and $1$. At commensurate densities, say, $\rho = 1/2$,
the charge density wave nature of the phase is clearly observed  for
$V = 5.6$ in Fig.~\ref{fig:fig3}(c). Alternative variation of the
density of bosons between one and zero is the signature of CDW-I
phase, $\mid 1~0~1~0~1~0~\cdots \ra$ type~\cite{kuhner,batrouniprl}.
Similarly, for density $\rho=1$, the density oscillation of the type
$\mid 2~0~2~0~2~0~\cdots \ra$, for $V=9$ suggest the CDW-II phase.
From the gap, the compressibility and the density oscillations, we
can conclude that for $U=10$ and $\rho=1/2$, we have a SF to CDW-I
phase transition at $V \sim 2.8$. However, for $\rho=1$, there is no
superfluid phase and the transition is from MI to CDW-II at the
critical value $V_C\sim 5.4$. These results are consistent with the
earlier results in the literature~\cite{pai,kuhner,batrouniprl}.

We now turn our attention to the case when $\rho$ is not
commensurate to the lattice length, i.e., $\rho\ne 1/2$ or $1$. From
Fig.~\ref{fig:fig2}, we observe that the compressibility is always
finite for incommensurate densities indicating that these regions of
the phase diagram correspond to the superfluid phase. However, the
local density distribution and $\ocdw$ reveal the richness of the
phases present in the compressible regions of the phase diagram.
Interesting phases appear when the nearest neighbor interaction is
large enough to obtain a CDW-I or CDW-II phase at commensurate
densities. Let us first consider densities close to $1/2$. When $V$
is less than the critical value, $V_C \sim 2.8$, for the SF-CDW
transition, we expect only the superfluid phase. However, for $V >
V_C$ the ground state shows solitonic behavior for densities close
to $\rho=1/2$. Fig.~\ref{fig:fig3} shows the the local densities
$\langle n_i \rangle$ as a function of the lattice sites $i$ for
different densities. The panel labeled (c) corresponds to the
commensurate density $\rho=1/2$ where we clearly observe the CDW
nature of the ground state, (b) and (d) shows the density variations
of the ground state where one boson has been added and removed from
the system at $\rho = 1/2$ respectively. Similarly, panels (a) and
(e) show the density variations when two bosons have been added and
removed respectively. The density profiles can be understood as
follows. Moving away from commensurate densities, the solitons
distort the periodic ground state by breaking the long range
crystalline order as a modulation in the density wave that minimizes
the ground state energy of the system~\cite{batrouniprl,sondhi}.

\begin{figure}[ht]
  \centering
\includegraphics[width = 3.4in, angle = 0, clip = true]
{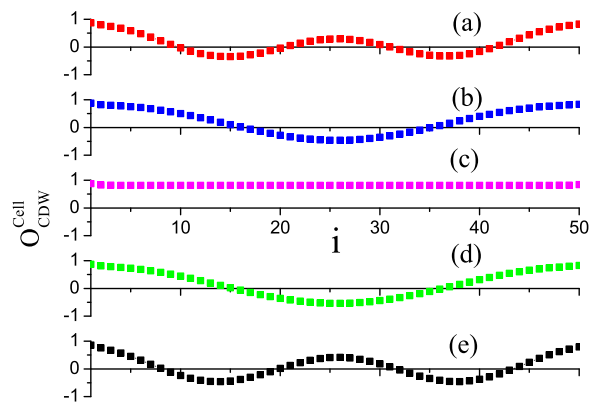}
    \caption{(Color online)CDW order paramater per unit cell, $\ocellcdw$ as a function of sites.
    (a) and (b) exhibits the solitonic signature
     for densities $\rho<1/2$. In panel (c), $\ocellcdw$ is flat,
     thereby indicating CDW-I with each alternate site occupied with one boson.
     (d) and (e) show the signature for the solitons at densities $\rho>1/2$.}
    \label{fig:fig4}
\end{figure}

\begin{figure}[ht]
  \centering
\includegraphics[width = 3.4in, angle = 0, clip = true]
{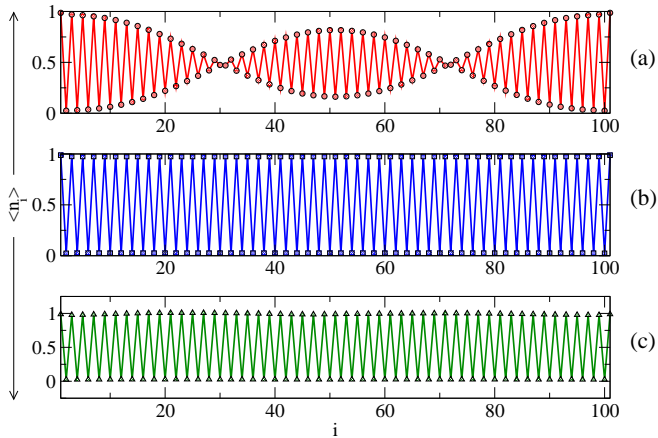}
    \caption{(Color online)Variation of the local density $\la n_i \ra$ as a function of $i$ for $V= 9$.
     Panel (a) shows the ground state density when a boson is removed from the state at $\rho=1/2$
     shown in panel (b), which is a CDW-I and panel (c) corresponds to the state obtained by adding
     a boson to $\rho=1/2$ state.}
     \label{fig:fig5}
\end{figure}

To understand the solitons we calculate the CDW order parameter for
each unit cell. In contrast to the superfluid and Mott insulator
phases that have one site per unit, the unit cell of CDW phase
consists of two lattice sites. Referring to these two sites as $1$ and
$2$, we define the CDW order parameter per unit cell as \beq
\ocellcdw=\langle n_1\rangle- \langle n_2 \rangle.
 \label{eq:Ocell}
\eeq The CDW phase has two degenerate ground states corresponding to
the two local density distributions $\mid1~0~1~0~1~0~ \cdots \rangle$
and $\mid0~1~0~1~0~1~\cdots \rangle$. The CDW order parameter,
$\ocellcdw$, for these two degenerate states is equal to $1$ and $-1$
respectively.
 Figure ~\ref{fig:fig4} shows the $\ocellcdw$ for the same set of densities
 considered in Fig.~\ref{fig:fig3}.
In Fig.~\ref{fig:fig4}, the center panel (c) has density $\rho=1/2$,
while (b) and (d) represent the system in panel (c) with one boson
added and removed respectively and (a) and (e) have two bosons added
and removed with respect to (c). We notice that the $\ocellcdw$ is
 uniform and close to one for $\rho=1/2$. Since we work with odd number of sites with
open boundaries, energy consideration leads to a CDW ground state which
represents $\mid1~0~1~0~1~0~ \cdots \rangle$ state. When we add or
remove one boson from this state, we get two solitons that modulate
the density distribution and break the long-range crystalline order.
The extra particle or hole splits into two solitons of equal
mass~\cite{sondhi,ramsesha}. The two solitons can move across the
lattice without causing any energy, however, if we want to get rid
of them, we need to spend lots of energy to flip the bosons.

\begin{figure}[ht]
  \centering
\includegraphics[width = 3.4in, angle = 0, clip = true]
{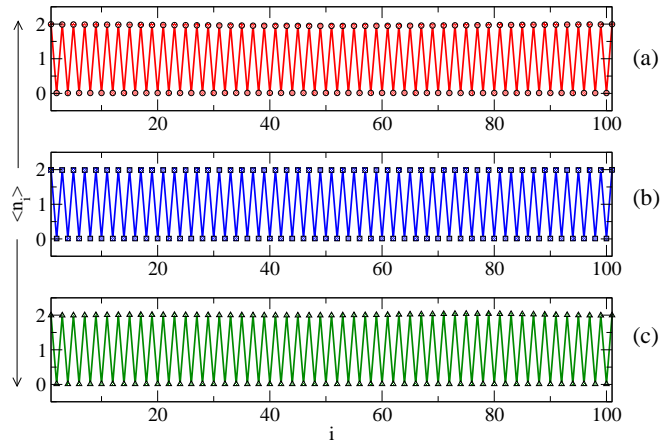}
    \caption{(Color online)Variation of the local density as a function of the lattice sites $i$ for doping around $\rho = 1$. Panel (b) corresponds to the CDW-II phase at $\rho = 1$ and panels
   (a) and (c) to the state obtained by removing and adding a boson to the CDW-II state. Notice that the crystalline structure is preserved for these density changes. $V =9$}
     \label{fig:fig6}
\end{figure}

Similarly, removal or addition of two bosons result in four
solitons. Continuing this process results in more solitons, until a
critical density is reached, when the density oscillation completely
dies out and the superfluid phase is obtained. Therefore, starting
with the CDW-I phase and changing the density from its commensurate
value of $\rho = 1/2$ by either adding or removing bosons, leads to
solitons+SF phase that finally becomes a superfluid. The transition
from solitonic to superfluid is more like a \emph{crossover} rather
than a real phase transition. The solitonic phase is obtained only
very close to $\rho=1/2$ and remains stable for the entire range of
$V$ considered on the hole side ($\rho < 1/2$). However, on the
particle side ($\rho > 1/2$), the solitonic phase remains stable
only up to some critical value of $V=V_C\sim 6.4$. For $V > 6.4$,
doping below half-filling breaks the CDW ground state into a
solitonic state that eventually goes into a superfluid phase.
However, this does not happen when bosons are added above
half-filling. For example, the variation of $\langle n_i \rangle$ as
a function of sites $i$ for three different densities are given in
Fig.~\ref{fig:fig5}, for $V=9$. The panel (b) represents the CDW-I
phase at $\rho = 1/2$ while (a) and (c) respectively, correspond to
the ground state with density
 obtained by removing
and adding one boson to the CDW-I ground state. While a solitonic
phase appears when bosons are removed ($\rho < 1/2$),
Fig.~\ref{fig:fig5}(c) suggest that the CDW-I phase is robust for
$\rho > 1/2$. Similar behavior is also seen when doping around
$\rho=1$. Fig.~\ref{fig:fig6} shows $\langle n_i \rangle$ as a
function of $i$ for densities around $\rho = 1$. Panels (a) and (c)
correspond to the density of the ground state obtained by removing
and adding one boson to the CDW-II state (panel (b)) at a density
$\rho = 1$.

\begin{figure}[ht]
  \centering
\includegraphics[width = 3.4in, angle = 0, clip = true]
{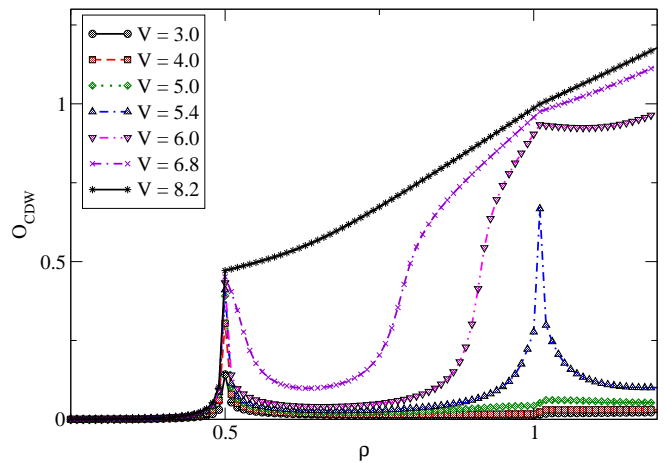}
    \caption{(Color online)$\ocdw$ as a function $\rho$ for different $V$. The two peaks at commensurate densities shows the existence of CDW-I and the CDW-II phases. The finite order parameter for large values of $V$ in the incommensurate density range shows the signature of the supersolid phase.}
     \label{fig:fig7}
\end{figure}

It turns out that for large $V$, the region between $1/2 < \rho <
1$, i.e., between CDW-I and CDW-II,  always remains in the CDW phase
even though the density is not commensurate to the lattice length.
The CDW order in the system is determined by calculating the CDW or-
der parameter given by Eq.~\ref{eq:ocdw} and is given in
Fig.~\ref{fig:fig7}. For small values of $V$, the $O_{CDW}$ is zero
for all the densities except at $\rho=1/2$ signalling the CDW-I
phase. However, as $V$ increases, an additional peak develops at
$\rho=1$ for $V > 5.4$ which corresponds to the CDW-II phase. The
most interesting feature is that the $O_{CDW}$ remains finite in the
region $1/2 < \rho < 1$ for large values of $V$. It may be noted
that the compressibility for $1/2 < \rho < 1$ is always finite. So
the bosons move freely on the CDW background and prefer to occupy
sites which are already occupied. Consider a system with one extra
boson added to the ground state corresponds to CDW-I. If the added
boson occupies an empty site, the energy cost is only due to the
nearest neighbor interaction and is of the order of $2V$. However,
if the added boson occupies a site which is already occupied by an
another boson, the energy cost is due to the onsite interaction and
is of the order of $U$ which is relatively smaller than $2V$ for
large $V$. The extra bosons therefore move between the occupied
sites with a finite hopping amplitude leading to a long range
correlation in the lattice, which yields to finite compressibility
in the region $1/2 < \rho < 1$ as shown in Fig.~\ref{fig:fig2}.
Similar behavior persists for doping above $\rho> 1$ as given in
Fig.~\ref{fig:fig6}.

\begin{figure}[ht]
  \centering
\includegraphics[width = 3.4in, angle = 0, clip = true]
{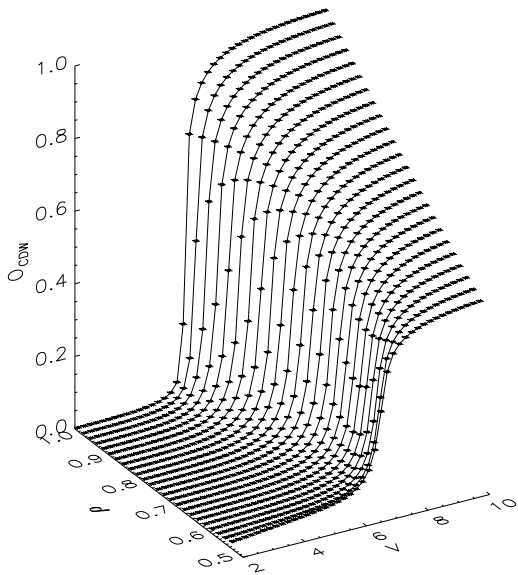}
    \caption{The CDW order parameter $\ocdw$ as a function of $V$ for different densities in the range
    $1/2 \le \rho \le 1$. Note that $\ocdw$ increases as $V$ increases.}
     \label{fig:fig8}
\end{figure}

Therefore we can conclude that for small $V$ apart from commensurate
fillings there exists no finite CDW order that is gapless and
incompressible. But for large $V$ the CDW order remains finite for
incommensurate densities. As a result, the region in the phase
diagram between CDW-I and CDW-II exhibits the coexistence of both
the diagonal long range order(DLRO) and the off-diagonal long range
order(ODLRO) which is the signature of the supersolid. In order to
obtain the boundary that separates the supersolid phase in the phase
diagram, we plot the $\ocdw$ with respect to $V$ for different
densities as seen in Fig.~\ref{fig:fig7}. We note that the $\ocdw$
increases sharply at some critical value of $V$, highlighting the
transition to the CDW phase. To obtain this critical value of $V$,
we take the derivatives of  $\ocdw$ with respect to $V$ for
different densities in the  range $1/2<\rho<1$ and $\rho >1$. The
point where the derivative is a maximum is taken to be  the critical
point of transition to the CDW phase. The order parameters $\ocdw$
as well as their derivatives as a function of $V$ are shown in
Fig.~\ref{fig:fig8}. The derivative shows a negligible peak for
$\rho<1/2$, but it shows a sharp peak for $\rho>1/2$ indicating the
existence of the CDW phase.

\begin{figure}[ht]
  \centering
\includegraphics[width = 3.4in, angle = 0, clip = true]
{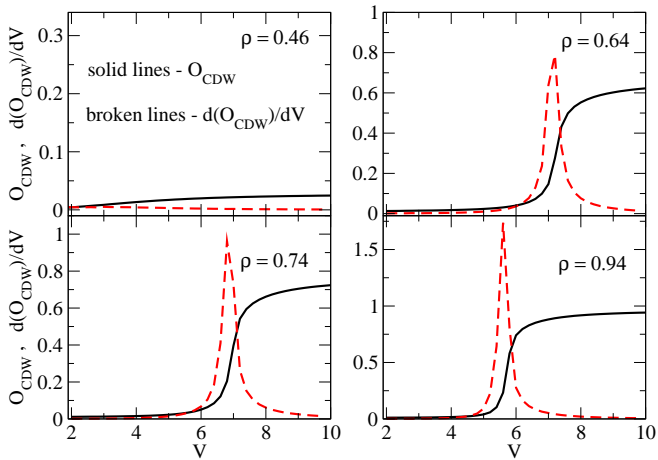}
  \caption{(Color online)CDW order parameters $\ocdw$ (solid lines) and their derivatives with respect to
  $V$ (broken lines) are given. For $V \ge V_C \sim 6.4$, the derivatives of the order
  parameters show peak at the transition to the CDW phase.
  }
     \label{fig:fig9}
\end{figure}

\begin{figure}[ht]
  \centering
\includegraphics[width = 3.4in, angle = 0, clip = true]
{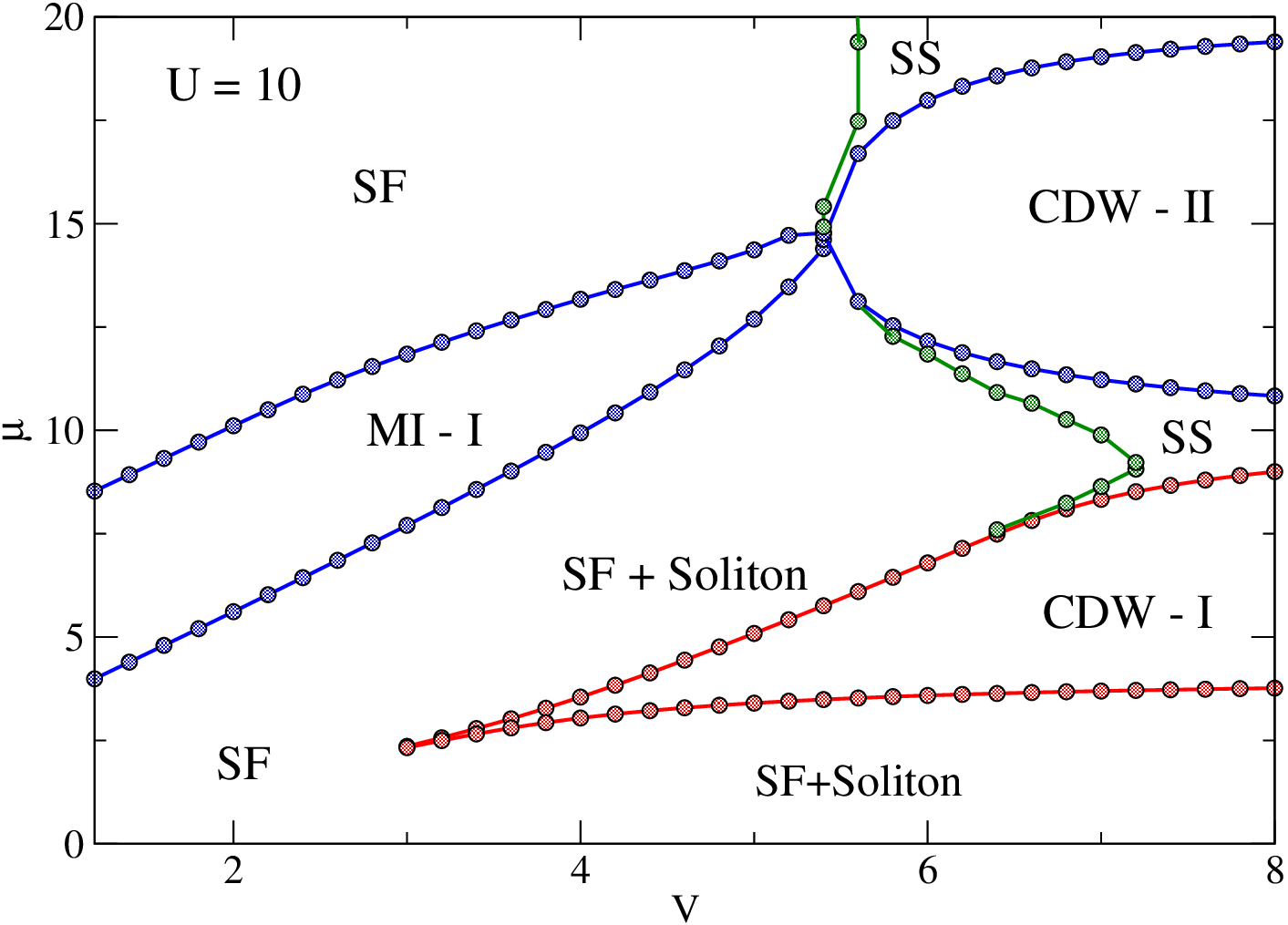}
  \caption{(Color online)Phase diagram showing all the phases for $U=10$ }
     \label{fig:fig10}
     \includegraphics[width = 3.4in, angle = 0, clip = true]
{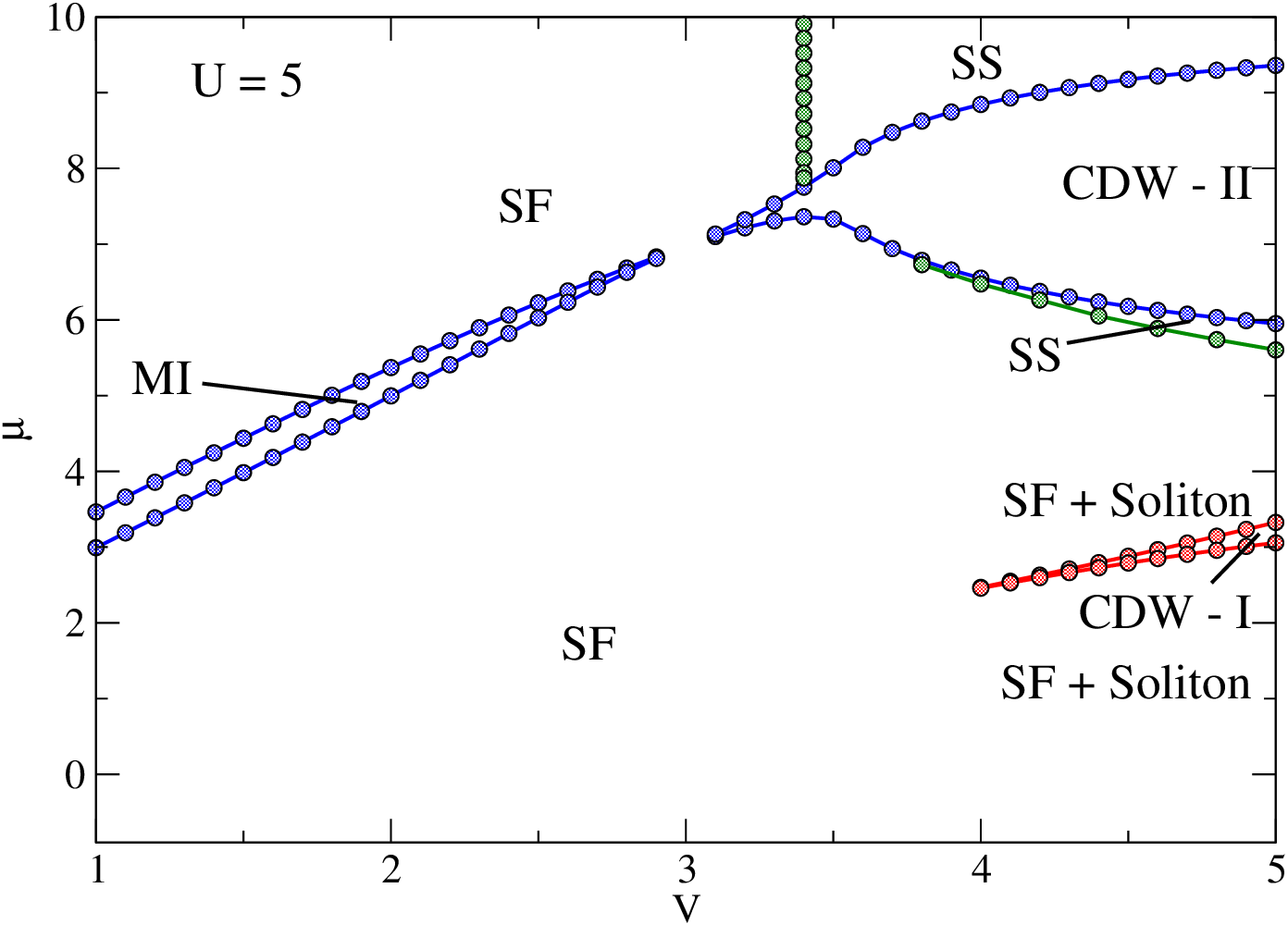}
  \caption{(Color online)Phase diagram showing all the phases for $U=5$ }
     \label{fig:fig11}
\end{figure}

The phase diagram obtained by plotting the chemical potential,
$\mu$, corresponding to different densities as a function of $V$ are
given in Fig.~\ref{fig:fig10}. To identify the region where the
gapped phases exist, we calculate the chemical potentials $\mu^+$
and $\mu^-$ ~\cite{whiteprb,ramanan} at $\rho=1/2$ and $1$ for all
values of $V$ in the thermodynamic limit and plot them in the $\mu -
V$ plane. The boundary of the supersolid phase is obtained by
calculating the chemical potential $\mu$ for $0.5<\rho<1$ and
$\rho>1$ at the critical value of $V$ where the system enters into
the the CDW phase. The critical value of $V$ for the transition to
the SS phase for density close to $\rho = 0.5$ is $V_C \sim 6.4$.
This critical value depends on density, exhibiting an increase as
the system is further doped and has a minimum value of $V_C = 5.4$.
For large values of V , it is clearly seen that the system continues
to be in the supersolid phase above CDW-II, while for smaller values
$V$ , it is in the SF phase.

The results remain qualitatively similar when $U=5$ and the
corresponding phase diagram is given in Fig.~\ref{fig:fig11}. In
this case the gapped regions such as CDW-I, CDW-II and MI shrink. There
is no direct transition from MI to CDW-II in sharp contrast to $U=10$.
Rather there are continuous MI-SF and SF-CDW-II transitions. The
supersolid phase occurs in a small region close to $\rho \lesssim
1$, but the trend is similar to that of $U=10$ for $\rho
>1$.

%%%%%%%%%%%%%%%%%%%%%%%%%%%%%%%%%%%%%%%%%%%%%%%%%%%%%%%%%%%%%%%%%%
\section{CONCLUSIONS}
\label{sect:concl} In summary, we have obtained the complete phase
diagram for a single species bosonic atoms in the framework of the
extended Bose-Hubbard model for two different values of the onsite
interaction $U$. Our studies have been carried out using the FS-DMRG method for
a large range of densities;  $0.25 \le \rho \le 1.25$. In
the large $U$ limit, we obtain CDW-I, MI, CDW-II, SF, soliton and
the supersolid phases and the transitions between them occurring at
various critical values of the nearest neighbor interaction. The
supersolid phase appears in the density range $0.5<\rho<1$ and
$\rho>1$ only in the large $V$ regime. The solitons are found to
exist for doping above half filling in the small $V$ regime and for
doping below half filling for the entire
 range of $V$. For an onsite interaction of intermediate strength($U=5$),
we find an interesting change in the phase diagram. The supersolid
phase becomes very small in the density range $0.5<\rho<1$ and it
exists only at densities close to $1$.

From an experimental point of view, in addition to the optical lattice, there is always
a magnetic trap present, thereby making these systems inhomogeneous.
Therefore, it makes it imperative to study this model in the
presence of a harmonic trap, where all the phases coexist.
Hence it becomes important to look for experimental signatures of
these phases in the presence of a trap. We are currently working
in that direction.

%%%%%%%%%%%%%%%%%%%%%%%%%%%%%%%%%%%%%%%%%%%%%%%%%%%%%%%%%%%%%%%%%%%

\section{ACKNOWLEDGMENT}
We thank G.~Baskaran and D.~Sen for useful discussions and comments.
RVP would like to thank DST and CSIR (India) for support and C. N.
Kumar for useful discussions. SR thanks Markus M\"{u}ller for useful
discussions.
%%%%%%%%%%%%%%%%%%%%%%%%%%%%%%%%%%%%%%%%%%%%%%%%%%%%%%%%%%%%%%%%%%%%

\end{document}